\documentclass[11pt]{article}
\newcommand{\Comment}[1]{{}}
\usepackage{amsfonts,amsthm,amsmath,amssymb,slashed}
\usepackage[textwidth = 430 pt, textheight = 630 pt]{geometry}

\usepackage{indentfirst}

\usepackage{hypernat}
\usepackage{graphicx}
\usepackage{cite}
\usepackage{siunitx}
\usepackage[font=small]{caption}

\usepackage{color}
\definecolor{MyDarkBlue}{rgb}{0.15,0.15,0.45}
\usepackage[linktocpage=true]{hyperref}
\hypersetup{
  colorlinks=true,
  citecolor=MyDarkBlue,
  linkcolor=MyDarkBlue,
  urlcolor=MyDarkBlue,
  pdfauthor={P Meert, H Nastase},
  pdftitle={Perturbation scalar DBI geometries},
  pdfsubject={hep-th}
}

\newcommand\ignore[1]{}
\def\one{{\,\hbox{1\kern-.8mm l}}}

\newcommand{\ie}{\emph{i.e.}\:}
\newcommand{\eg}{\emph{e.g.}\;}

\newcommand{\Cset}{{\,\,{{{^{_{\pmb{\mid}}}}\kern-.45em{\mathrm C}}}}}

\newcommand{\be}{\begin{equation}}
  \newcommand{\bea}{\begin{eqnarray}}

  \newcommand{\ee}{\end{equation}}
\newcommand{\eea}{\end{eqnarray}}

\newcommand{\el}{\ell}

\begin{document}

\renewcommand{\thefootnote}{\fnsymbol{footnote}}

\makeatletter
\@addtoreset{equation}{section}
\makeatother
\renewcommand{\theequation}{\thesection.\arabic{equation}}

\rightline{}
\rightline{}
%   \vspace{1.8truecm}

%\begin{flushright}
% preprint nrs.
%\end{flushright}

%\vspace{10pt}

%%%%%%%%%%%%%%%%%

\begin{center}
  {\LARGE \bf{\sc Static and Dynamic perturbations of  scalar DBI analogue of a black hole}}
\end{center}
\vspace{1truecm}
\thispagestyle{empty} \centerline{
  {\large \bf {\sc Pedro Meert${}^{a}$}}\footnote{E-mail address: \Comment{\href{mailto:pedro.meert@unesp.br}}
  {\tt pedro.meert@unesp.br}}
  {\bf{\sc and }}
  {\large \bf {\sc Horatiu Nastase${}^{a}$}}\footnote{E-mail address: \Comment{\href{mailto:horatiu.nastase@unesp.br}}
  {\tt horatiu.nastase@unesp.br}}
 % {\bf{\sc and }}
  %{\large \bf {\sc Justin Khoury${}^{b}$}}\footnote{E-mail address: \Comment{\href{mailto:justing@email.com}}
  %{\tt justin@email.com}}
}

\vspace{.5cm}

%\vspace{.3cm}

\centerline{{\it ${}^a$Instituto de F\'{i}sica Te\'{o}rica, UNESP-Universidade Estadual Paulista}}
\centerline{{\it R. Dr. Bento T. Ferraz 271, Bl. II, S\~ao Paulo 01140-070, SP, Brazil}}
%TODO: Justin affiliation here

\vspace{1truecm}

%%%%%%%%%%%%%%%%%
\thispagestyle{empty}

\centerline{\sc Abstract}

\vspace{.4truecm}

\begin{center}
  \begin{minipage}[c]{380pt}
    {\noindent
      Perturbations around a spherically symmetric solution of the scalar Dirac-Born-Infeld (DBI) theory obey the
      equation of motion of a massless scalar field in a curved background with black hole-like properties
      like finite temperature and an event horizon with zero speed of propagation of information (a black hole 
      analogue). 
      After reviewing some properties of
      the background geometry for the propagations, we investigate the solutions considering static and
      dynamic regimes. In the former case we obtain the so-called Love numbers, or more generally scalar
      polarizability. In the latter case, we employ boundary conditions analogous to the computation of
      quasi-normal modes, and argue that the background geometry acts as a classical high-pass filter.
      We therefore provide further evidence that the scalar black hole analogues can be analyzed similarly 
      to a black hole.
    }
  \end{minipage}
\end{center}

\vspace{.5cm}

\setcounter{page}{0}
\setcounter{tocdepth}{2}

\newpage

\tableofcontents
\renewcommand{\thefootnote}{\arabic{footnote}}
\setcounter{footnote}{0}

\linespread{1.1}
\parskip 4pt

%% Text start

\section{Introduction}\label{Sec:Intro}

%Include more context about the scalar DBI model.

Following the discovery of Unruh that there exist fluid flows that mimic the effects of black holes
(what he called "dumb holes", {\em i.e.}, no-sound holes) \cite{Unruh:1980cg,Unruh:1994je}, the idea of 
black hole analogues has slowly gained ground, both in fluid flows and, more recently, in condensed matter 
systems, and has become a subject of experimental research (for a review of analogue gravity, see
\cite{Barcelo:2026vlr}). On the other hand, the AdS/CFT correspondence started by Maldacena
\cite{Maldacena:1997re} (see the books \cite{Nastase:2015wjb,Ammon:2015wua} for an introduction, and 
\cite{Nastase:2017cxp} for its applications to condensed matter physics)
relates gravity in a bulk geometry to a nonperturbative field theory on its boundary (and a nonperturbative 
field theory is describable at large distances as a fluid expansion in derivatives), as well as relating 
the near-horizon physics of a black hole in the bulk to a fluid, via the "fluid-gravity correspondence"
\cite{Bhattacharyya:2007vjd}. This already hints to the fact that there could be a field theory analogue of a 
black hole, perhaps via an intermediate step of a fluid analogue. 

In a related development, the fireballs of strongly-coupled quark-gluon (sQGP) plasma 
obtained in high energy heavy ion collisions at RHIC and ALICE
were described in terms of a black hole in the gravity dual bulk, one which, however, is pancake-shaped, 
and effectively lives on the IR cut-off of the bulk, becoming effectively 3+1 dimensional
\cite{Nastase:2005rp,Aharony:2005bm,Nastase:2008hw}. But it turns out that in field theory, this (asymptotic
at high energy) collision with fireball creation is well described by the very simple model introduced by Heisenberg
of a scalar with a DBI action \cite{Kang:2004jd,Kang:2005bj,Nastase:2015ljb}. 

Then it becomes obvious that one should ask if the DBI scalar model can admit a field theory analogue of a 
black hole, that should describe the sQGP fireball and be dual to the black hole in the bulk. Following  
an only partially successful attempt in \cite{Nastase:2005pb}, it was shown in \cite{Nastase:2007ut}
that the answer is yes, and moreover, there is a {\em unique} scalar action that admits such a solution, 
of the type of DBI plus a particular source term, standing in for the colliding heavy ions. 

In~\cite{Nastase:2007ut}, it was noted that the equations governing perturbations around solutions of the
scalar DBI model can be mapped to a massless scalar field propagating on a curved background. One can thus
leverage methods typically used in analogue gravity models (\eg,~\cite{Barcelo:2026vlr}) to study these
perturbations. We show that, depending on the source of the DBI solution, one can associate different
effective geometries for the background. This “field-fluid” map allows us to read off a metric function that
inherits the properties of the source, which is then used to study the behavior of the perturbations.

A particularly interesting aspect of static perturbations in standard General Relativity (GR) is the vanishing
of Love numbers for black holes~\cite{Binnington:2009bb,Damour:2009va}. This result holds for all
asymptotically flat, vacuum black holes in 4 dimensions and is a striking feature of GR perturbation theory.
It is typically associated with the lack of internal structure, as these numbers quantify the tidal
deformability of an object in response to an external field. 

Inspired by recent investigations of the tidal
properties of analogue black holes~\cite{DeLuca:2024nih,DeLuca:2025zqr}, 
we compute the Love numbers associated with the
scalar DBI geometry and find a remarkable similarity. Namely, the point-like source exhibits two ladder
structures of vanishing Love numbers, whereas a smeared source model exhibits a logarithmically running Love
number.\footnote{Note that the Love numbers and tidal properties of solutions of DBI scalars were also 
considered in \cite{BeltranJimenez:2022hvs,BeltranJimenez:2024zmd,BeltranJimenez:2025gkx}, 
though the authors focused on the wrong-sign DBI model, not the one relevant for us.}

Dynamic perturbations are equally important, as they reveal intrinsic properties of the underlying geometry
regardless of the specific nature of the initial perturbation. In the context of black holes, these are
encapsulated by quasi-normal modes (QNMs). These modes are characteristic of dissipative spacetimes,
reflecting the purely ingoing boundary conditions imposed at the horizon and outgoing conditions at spatial
infinity. While a wide range of methods for computing such modes are available (see
\eg,~\cite{Kokkotas:1999bd,Berti:2009kk} for reviews), we argue that in the scalar DBI model, only
perturbations in the background of the smeared source allow for purely dissipative boundary conditions.
Adapting existing numerical methods, we show that this geometry does not possess a standard QNM spectrum.
Instead, it features sub-threshold virtual states (evanescent waves induced by the monotonically decreasing
potential of the background geometry). We conclude that this system acts as a classical high-pass filter, as
waves below a certain threshold energy are perfectly reflected, while those above it are perfectly absorbed
without ringing.

In this work we explore the static and dynamic perturbations of the scalar DBI model from the analogue gravity
perspective, extracting the so-called Love numbers and complex frequencies that, as we discuss, correspond to
evanescent waves reflected-off the source. In section~\ref{Sec:model} we review the results
of~\cite{Nastase:2007ut} that are relevant to the perturbation theory, and also discuss some details that were
previously overlooked. Section~\ref{Sec:LNumbers} contains the discussion of static perturbations in the
presence of point-like and smeared sources. In Sect.~\ref{Sec:QNMs} we explore dynamical perturbations on the
analogue geometry of the smeared source, after arguing that such perturbations are irrelevant in the context
of the point-like source. We conclude with a discussion of the results, comparison with recent findings and
future directions for this work.

\section{Model}\label{Sec:model}

In this section we review the relevant results derived in~\cite{Nastase:2007ut}, constructing the metric in a
“black hole-like” form. The scalar DBI model in 4 spacetime dimensions is described by the action
\begin{equation}
  S = b^{-2}\int d^{4}x\left[\sqrt{1+b^{2}\left(\partial\phi\right)^{2}}-1\right]+S_{\text{source}}
  \ ,
  \label{eq:DBIaction}
\end{equation}
where $ b $ is a parameter set to one throughout this work. Two source terms that will be considered are
\begin{align}
  S_{\text{source}} & =\int d^{4}x\phi C\delta\left(r\right) \label{eq:deltaSource}\ , \\
  S_{\text{source}} & =\int d^{4}x\phi\frac{\alpha}{r^{2}} \label{eq:smearedSource}\ ,
\end{align}
with $ C $ and $ \alpha $ coupling constants. The model with delta function source term was considered 
initially in \cite{Nastase:2005pb}, though one did not quite obtain the needed properties of a black hole. 
By contrast, the model with $\alpha$ was introduced in \cite{Nastase:2007ut} and was shown to be 
the unique scalar field model generating a black hole analogue solution with a horizon, 
with finite temperature
and vanishing propagation speed of information at the horizon. 
We consider a radial profile for the field $ \phi $, then the
equation of motion reads (putting $b=1$, as we said)
\begin{equation}
  \frac{r^{2}\phi^{\prime}}{\sqrt{1+\left(\phi^{\prime}\right)^{2}}}
  =\frac{\partial}{\partial\phi}\mathcal{L}_{\text{source}}
  \ ,
  \label{eq:genEOM}
\end{equation}
the Lagrangian $ \mathcal{L}_{\text{source}} $ depends on source given 
by Eqs.~\eqref{eq:deltaSource} and
\eqref{eq:smearedSource}. For the particular choice of radial profile, 
the catenoid solution can be written in
a single equation for $ \phi\left( r \right) $
\begin{equation}
  \phi\left(r\right)=\int_{r}^{\infty}\frac{\alpha x+C}{\sqrt{x^{4}-\left(\alpha x+C\right)^{2}}}
  \ .
  \label{eq:solGen}
\end{equation}
When $ \alpha=0 $ we have source given by~\eqref{eq:deltaSource}, 
while for source~\eqref{eq:smearedSource}
both $ \alpha $ and $ C $ are non-vanishing, in which case $ C $ is an integration constant.

The singular points of $ \phi $ are obtained from the function
\begin{equation}
  g_{\alpha}\left(r\right)=r^{4}-\left(\alpha r+C\right)^{2}
  \ ,
  \label{eq:gGen}
\end{equation}
where we use the label $ \alpha $ as follows: for $ \alpha = 0 $ the function corresponds to the Dirac delta
source~\eqref{eq:deltaSource}, when $ \alpha\neq 0 $ it corresponds to the smeared
source~\eqref{eq:smearedSource}. When $ \alpha = 0 $ it is clear from~\eqref{eq:solGen} 
that $ r_0^{2} = C $
is such that $ \phi^{\prime}\to\infty $ while $ \phi $ remains finite. For general $ \alpha $ we are
interested in the case when $ g_{\alpha}\left(r\to r_{0}\right)\approx A\left(r-r_{0}\right)^{2} $. There are
two solutions that give the second order approximation
\begin{equation}
  C=-\frac{\alpha^{2}}{4},\text{ } r_{0}=\frac{\alpha}{2}\ ,
  \label{eq:couplings}
\end{equation}
and the same solutions with opposite signs - throughout this work we will 
use only~\eqref{eq:couplings}. This
choice for the couplings happens only for the smeared source, as it produces t
he following behaviour for the field:
\begin{align}
  \begin{aligned}
    \phi^{\prime}\left( r\to r_0 \right) & \approx       \frac{r_{0}}{\sqrt{2}\left(r-r_{0}\right)}\ ,        \\
    \phi\left(r \to r_0 \right)          & \approx  \phi_{0}+\frac{r_{0}}{\sqrt{2}}\ln\left(r-r_{0}\right)\ .
    \label{eq:smearedSolApprox}
  \end{aligned}
\end{align}

Consider the fluctuations $\phi=\phi_{0}\left(r\right)+\delta\Phi$ where $\phi_{0}\left(r\right)$ is the
solution~\eqref{eq:solGen}. The equation of motion for $\delta\Phi$ reads
\begin{equation}
  -\partial_{t}\left[\frac{\partial_{t}\delta\Phi}{\sqrt{1+\left(\nabla\phi_{0}\right)^{2}}}\right]
  +\partial_{i}
  \left[
    \frac{1}{\sqrt{1+\left(\nabla\phi_{0}\right)^{2}}}
    \left(
      \delta^{ij}
      -
      \frac{\partial^{i}\phi_{0}}{\sqrt{1+\left(\nabla\phi_{0}\right)^{2}}}
      \frac{\partial^{j}\phi_{0}}{\sqrt{1+\left(\nabla\phi_{0}\right)^{2}}}
    \right)\partial_{j}\delta\Phi
  \right]
  =
  0
  \ ,
  \label{eq:DBIpert}
\end{equation}
which is the equation for a massless scalar field in a curved background. It was shown
in~\cite{Nastase:2007ut} that a using the relations
\begin{align}
  c^{2}_{s} & =1+\left(\nabla\phi_{0}\right)^{2} \label{eq:cs} ,                    \\
  \rho_{0}  & =\frac{1}{\sqrt{1+\left(\nabla\phi_{0}\right)^{2}}} \label{eq:rho0} , \\
  v^{i}_{0} & =\partial^{i}\phi_{0} \label{eq:v0} ,
\end{align}
Eq.~\eqref{eq:DBIpert} reads
\begin{equation}
  -\partial_{t}
  \left[\frac{\rho}{c^{2}_{s}}
    \left(
      \partial_{t}\phi_{0}
      +
      \vec{v}_{0}\cdot\nabla\phi_{0}
    \right)
  \right]
  +
  \nabla\cdot
  \left[
    \rho_{0}\nabla\phi_{0}-\frac{\rho\vec{v}_{0}}{c^{2}_{s}}
    \left(\partial_{t}\phi_{0}+\vec{v}_{0}\cdot\nabla\phi_{0}\right)
  \right]=0
  ,
\end{equation}
which is the equation for the fluctuations of an irrotational, barotropic inviscid fluid~\cite{Unruh:1980cg}.
The curved background is referred to as acoustic metric, the line element reads
\begin{equation}
  ds^{2}=\frac{\rho_{0}}{c_{s}}
  \left[
    -c^{2}_{s}dt^{2}
    +
    \delta_{ij}\left(dx^{i}-v^{i}_{0}dt\right)\left(dx^{j}-v^{j}_{0}dt\right)
  \right]
  .
  \label{eq:acousticmetric}
\end{equation}
We can identify a ``black hole like'' metric after the coordinate transformation
$dt=d\tau-\frac{v^{i}_{0}dx^{i}}{c^{2}_{s}-v^{2}_{0}}$ and using spherical coordinates
\begin{equation}
  ds^{2}=\frac{\rho_{0}}{c_{s}}
  \left[
    -\left(1-\frac{v^{2}_{0}}{c^{2}_{s}}\right)c^{2}_{s}d\tau^{2}
    +
    \frac{dr^{2}}{\left(1-\frac{v^{2}_{0}}{c^{2}_{s}}\right)}
    +
    r^{2}d\Omega_2^{2}
  \right]
  .
  \label{eq:acousticbhlike}
\end{equation}
For the particular case of the scalar DBI map, Eqs.~\eqref{eq:cs} -~\eqref{eq:v0}, 
the following relations hold:
\begin{align}
  c^{2}_{s}-v^{2}_{0} & =1 \\
  \rho_{0}c_{s}       & =1
\end{align}
These allow to further simplify Eq.~\eqref{eq:acousticbhlike} to
\begin{equation}
  ds^{2}=\frac{1}{c^{2}_{s}}\left[-d\tau^{2}+c^{2}_{s}dr^{2}+r^{2}d\Omega_2^{2}\right]
  .
  \label{eq:acousticDBImetric}
\end{equation}
Eq.~\eqref{eq:acousticDBImetric} is a direct consequence of the relations between the fluid variables and
scalar DBI solution. From~\eqref{eq:cs} and the solution~\eqref{eq:solGen} 
we can write the metric in terms of
$ r $ only.

A simple calculation reveals that
\begin{equation}
  \frac{1}{c^{2}_{s}}
  =
  1-\frac{\left(\phi^{\prime}\right)^{2}}{1+\left(\phi^{\prime}\right)^{2}}
  =
  1-\left(\frac{\alpha}{r}+\frac{C}{r^{2}}\right)^{2}=\frac{g_\alpha}{r^4}\equiv f_{\alpha}\;,
\end{equation}
where we can now use the couplings for the sources and their relations to $r_{0}$ to write the line
element~\eqref{eq:acousticDBImetric} in terms of a function we will call
$f_{\alpha}=f_{\alpha}\left(r\right)$. The delta source corresponds to $\alpha=0$, and as discussed
previously, $r^{2}_{0}=C$ in this case, such that
\begin{equation}
  f_{0}=1-\left(\frac{r_{0}}{r}\right)^{4}
  .
  \label{eq:f0}
\end{equation}
For the smeared source, we have $\alpha\neq0$, and 
the particular choice of couplings in Eq.~\eqref{eq:couplings}
yields
\begin{equation}
  f_{\alpha}=1-\left(\frac{r_{0}}{r}\right)^{4}\left(1-2\frac{r}{r_{0}}\right)^{2}
  .
  \label{eq:falpha}
\end{equation}
Expressions~\eqref{eq:f0} and~\eqref{eq:falpha} are just convenient notations for the analysis in the next
sections. Near the horizon, the function $ f_\alpha $ has the behavior
\begin{align}
  f_{0}\left(r_{0}\right)      & \approx\frac{4}{r_{0}}\left(r-r_{0}\right), \label{eq:f0nearr0}            \\
  f_{\alpha}\left(r_{0}\right) & \approx\frac{2}{r^{2}_{0}}\left(r-r_{0}\right)^{2} \label{eq:falphanearr0}
  .
\end{align}
For the metric~\eqref{eq:acousticDBImetric}, the tortoise coordinate is
\begin{equation}
  dr_{*}=\frac{dr}{\sqrt{f_{\alpha}}}
  ,
  \label{eq:tortoisedef}
\end{equation}
from which one finds that the tortoise coordinate associated with $\alpha=0$ behaves as
$r_{*}\sim\sqrt{r-r_{0}}$ for $r\to r_{0}$. This is in contrast with the logarithmic behavior of the smeared
source in the same regime. This will be crucial in Sect.~\ref{Sec:QNMs}, when we analyze dynamic
perturbations, as the near horizon behavior and the interval in which $r_{*}$ is defined determine the
boundary conditions for the problem. It should also be noted that the finite tortoise coordinate for the delta
source model is related to the diverging temperature and finite phase and group velocities of the waves at
$r_{0}$. In fact, this is the motivation for introducing the smeared source, which allows for the
interpretation of $r_{0}$ as a thermal horizon~\cite{Nastase:2007ut}.
Indeed, in this second case, since $r_*\simeq r+(\frac{\alpha}{2\sqrt{2}})\ln (r-r_0)$, similar to the case of 
a black hole, one obtained both phase and group velocities vanishing near the horizon, 
\begin{equation}
c_{\rm ph}=\frac{\omega}{k}\propto \sqrt{r-r_0}\rightarrow 0\;,\;\;
c_{\rm g}=\frac{d\omega}{dk}\propto \sqrt{r-r_0}\rightarrow 0\;,
\end{equation}
so we have indeed an event horizon for information, 
and a finite horizon temperature (note that $[\alpha]=1$ and $[b]=-2$)
\begin{equation}
T=\frac{\sqrt{2}}{\pi \alpha b}.
\end{equation}

\section{Static perturbations: Love numbers}\label{Sec:LNumbers}

% $ \frac{1}{\sqrt{-g}}\partial_{\mu}\left(\sqrt{-g}g^{\mu\nu}\partial_{\nu}\delta\Phi\right)=0 $

We are going to consider static perturbations $ \delta\Phi = \delta\Phi\left( r,\theta,\varphi \right) $ for
the equation of the massless scalar field in the background given by the metric~\eqref{eq:acousticDBImetric}.
Explicitly, the equation reads
\begin{equation}
  \frac{1}{r^{2}f_{\alpha}^{3/2}}\partial_{r}\left(r^{2}f_{\alpha}^{3/2}\partial_{r}\delta\Phi\right)
  +\frac{1}{r^{2}f_{\alpha}\sin\theta}\left[\partial_{\theta}\left(\partial_{\theta}\sin\theta\delta\Phi\right)
  +\frac{1}{\sin\theta}\partial_{\phi}^{2}\delta\Phi\right]
  =0
  \ ,
  \label{eq:fullstaticperturbationeq}
\end{equation}
such that the ansatz $ \delta\Phi=\frac{R\left(r\right)}{r}Y_{\el m}\left(\theta,\phi\right)
$ separates the equation. 

Before specializing to each source model, we discuss the general 
aspects that concern the calculation of the
scalar polarizability, which is the analogue of the gravitational tidal deformation, or Love number. Note that
in the asymptotic limit $ r\to \infty $ the function $ f_{\alpha}\to 1 $ faster than $1/r^2$, such
that the radial equation from~\eqref{eq:fullstaticperturbationeq} reads
\begin{equation}
  R^{\prime\prime}_{\el} - \frac{\el\left( \el +1 \right)}{r^{2}}R_{\el} = 0
  \ .
  \label{eq:inftystaticperturbationeq}
\end{equation}
The Frobenius method dictates the scaling of the solutions, such that at spatial infinity the perturbations
behave as
\begin{equation}
  \delta\Phi\left( r\to\infty \right) \sim
  \left( A r^{\el} + \frac{B}{r^{\el+1}} \right)Y_{\el m}\left( \theta,\phi \right)
  \ .
  \label{eq:inftystaticsol}
\end{equation}
In the context of linear response theory this asymptotic profile is decomposed into an externally applied
tidal field $ \varepsilon $ (the growing source mode), and the 
induced multipole moment $ \mathcal{M} $ (the
decaying response mode). The Love number is defined as the 
ratio between the induced response and the applied
source
\begin{equation}
  k_{\el} \propto \frac{\varepsilon}{\mathcal M} \propto \frac{B}{A}\ .
  \label{eq:Lovedef}
\end{equation}
In the following we discuss how to explicitly compute $ k_{\el} $ depending on each source.

\subsection{Point-like source}

In this case we consider $ \alpha=0 $, so $ f_{\alpha}=f_0 $ is given by~\eqref{eq:f0}. Using the
dimensionless coordinate
\begin{equation}
  z=\left(\frac{r_{0}}{r}\right)^{4}
  ,
  \label{eq:zdelta}
\end{equation}
the radial equation associated with~\eqref{eq:fullstaticperturbationeq} can 
be cast in the standard form of a
Gauss Hypergeometric equation
\begin{equation}
  \left(1-z\right)zR^{\prime\prime}
  +\left(\frac{5}{4}-\frac{11}{4}z\right)R^{\prime}-\left[\frac{3}{8}+\frac{\ell\left(\ell+1\right)}{16z}\right]R
  =0
  .
  \label{eq:hypergeomeq}
\end{equation}
The solutions of this equation will have the form $ R\left( z \right) = z^{p}\ _{2}F_{1}\left(a,b,c;z\right)$
where $ p $ is the scaling behavior of the asymptotic solution~\eqref{eq:inftystaticsol}, corrected by the
$ 1/4 $ factor due to the transformation~\eqref{eq:zdelta}, and $ a,b,c $ 
are coefficients determined from the
equation~\eqref{eq:hypergeomeq}.

The complete solution to the radial equation after substituting back the radial variable $ r $ reads
\begin{equation}
  \begin{split}
    R\left( r \right) = \
     & B \left(\frac{r_0}{r}\right)^{\el}
     \text{}_{2}F_{1}\left(\frac{\el+1}{4},\frac{\el+6}{4},\frac{2\el+5}{4};\frac{r_{0}^{4}}{r^{4}}\right) \\
     & \ \ \ +
     A \left( \frac{r}{r_{0}} \right)^{\el + 1}
     \text{}_{2}F_{1}\left(\frac{5-\el}{4},-\frac{\el}{4},\frac{3-2\el}{4};\frac{r_{0}^{4}}{r^{4}}\right)
     .
  \end{split}
  \label{eq:perturbsol}
\end{equation}
And one can easily check that upon substitution in the ansatz for $ \delta\Phi $, the field behavior at $ r\to
\infty $ is exactly that of~\eqref{eq:inftystaticsol}. To determine the values of $ A $ and $ B $ we
require the solutions be regular at the horizon $ r_0 $. 
The exact condition is\footnote{Using the transformation law 
for the hypergeometric function, that allows us to find the 
behaviour near $z=1$ from the behaviour near $z=0$,
\begin{eqnarray}
{}_2F_1(\alpha,\beta,\gamma;z)&=&\frac{\Gamma(\gamma)
\Gamma(\gamma-\alpha-\beta)}{\Gamma(\gamma
-\alpha)\Gamma(\gamma-\beta)}{}_2F_1(\alpha,\beta,\alpha+\beta-\gamma+1;1-z)\cr
&&+(1-z)^{\gamma-\alpha-\beta}\frac{\Gamma(\gamma)\Gamma(\alpha+\beta-\gamma)}{\Gamma(\alpha)
\Gamma(\beta)}{}_2F_1(\gamma-\alpha,\gamma-\beta,\gamma-\alpha-\beta;1-z).
\end{eqnarray}
}
\begin{equation}
  B
  \frac{\Gamma\left( \frac{2\el + 5}{4} \right)}{\Gamma\left( \frac{\el + 1}{4} \right)
  \Gamma\left( \frac{\el + 6}{4} \right)}
  +
  A
  \frac{\Gamma\left( \frac{3\el - 2}{4} \right)}{\Gamma\left( \frac{5 - \el}{4} \right)
  \Gamma\left(- \frac{\el}{4} \right)}
  =
  0
  \ ,
  \label{eq:deltaRegr0}
\end{equation}
obtained using properties of the hypergeometric functions~\eqref{eq:perturbsol} as $ r\to r_0 $. From
Eq.~\eqref{eq:Lovedef} the Love numbers are
\begin{equation}
  \kappa_{\el} = 
  \frac{B}{A} =
  -r_0^{2\el + 1}
  \frac{\Gamma\left(\frac{3-2\el}{4}\right)\Gamma\left(\frac{\el+1}{4}\right)\Gamma\left(\frac{6+\el}{4}\right)}
  {\Gamma\left(\frac{5+2\el}{4}\right)\Gamma\left(\frac{5-\el}{4}\right)\Gamma\left(-\frac{\el}{4}\right)}
  \ .
  \label{eq:DeltaLoveExpr}
\end{equation}
To determine when this expression vanishes we look at $ \Gamma $ function behavior. Considering that $ \el\geq 0
$ is an integer, the numerator will never vanish. On the other hand, $ \Gamma\left( x \right)\to \infty $ for
$ x = 0, -1, -2, -3,\ldots $, which happens for
\begin{equation}
  \el=4n+5 {\rm \ and \ } \el=4n\ ,
  \label{eq:gammainfty}
\end{equation}
for integer $ n $. So for $ \el=\left\{ 0,4,5,8,9,12,13,16,17,\ldots\right\} $ we have vanishing Love number.
This result is very similar to what has been previously found for the canonical analogue black hole
in~\cite{DeLuca:2024nih}.

\subsubsection*{Ladder Symmetry}

Following the approach of~\cite{Ghosh:2026vig}, we construct ladder operators that explain the periodic
vanishing of the Love numbers observed in~\eqref{eq:gammainfty}. Define the “Hamiltonian” operator multiplying
Eq.~\eqref{eq:hypergeomeq} by $ -\left(1-z\right)/z $:
\begin{equation}
  H_{\ell}
  =
  -\left(1-z\right)^{2}\frac{d^{2}}{dz^{2}}
  -\frac{\left(1-z\right)}{z}\frac{\left(5-11z\right)}{4}\frac{d}{dz}
  +\frac{\left(1-z\right)}{z}\left[\frac{3}{8}+\frac{\ell\left(\ell+1\right)}{16z}\right]
  ,
  \label{eq:LoveHamiltonian}
\end{equation}
so that the radial equation takes the form $ H_{\el}R = 0 $. This Hamiltonian can be factorized using the
first-order operators
\begin{align}
  D^{+}_{\ell}
   & =
   -\left(1-z\right)\frac{d}{dz}
   +\frac{\ell\left(1-z\right)}{4z}
   +\frac{\left(\ell+1\right)\left(\ell+6\right)}{4\left(2\ell+5\right)}
   ,
   \\
   D^{-}_{\ell}
   & =
   \left(1-z\right)\frac{d}{dz}
   +\frac{\left(\ell+1\right)\left(1-z\right)}{4z}
   +\frac{\ell\left(\ell-5\right)}{4\left(2\ell-3\right)}
   .
\end{align}
These ladder operators relate Hamiltonians of different multipoles in steps of $ \Delta\el=4 $
\begin{align}
  \begin{aligned}
    H_{\ell}&=D^{-}_{\ell+4}D^{+}_{\ell}+E_{\ell} ,
    \\
    H_{\ell}&=D^{+}_{\ell-4}D^{-}_{\ell}+E_{\ell-4}
    ,
  \end{aligned}
\end{align}
where the factorization constant $ E_{\el} $ is given by
\begin{equation}
  E_{\ell}=%-\frac{\ell^{4}+10\ell^{3}+23\ell^{2}-10\ell-24}{16\left(2\ell+5\right)^{2}}
  -\frac{(\ell^2-1)(\ell+4)(\ell+6)}{16(2\ell+5)^2}.
  \label{eq:Eell}
\end{equation}

We also obtain that 
\begin{equation}
H_{\ell\pm 4}D_\ell^\pm=D_\ell^\pm H_\ell
\end{equation}
when acting on $R_\ell$. 

As we will show, the values of $ \el $ such that $ E_{\el}=0 $ dictate the vanishing of Love numbers for the
values $ \el = 5,9,13\ldots $.

The key property of the ladder operators $ D^{\pm}_{\el} $ is that they do not mix source and response modes.
Specifically, the raising operator maps a solution at multipole $ \hat{\el} $  to a solution at $ \hat{\el}+4
$, \ie,  $ D^{+}_{\hat\el}R_{\hat{\el}} = R_{\hat{\el}+4} $. Asymptotically, $ R_{\hat\el} $ reads
\begin{equation}
  R_{\hat\ell}\left(z\right)\sim A_{\hat\ell}z^{-\frac{\hat\ell+1}{4}}+B_{\hat\ell}z^{\frac{\hat\ell}{4}}
  .
  \label{eq:LoveasympDelta}
\end{equation}
If $ B_{\hat\el} = 0 $, the raised solution will also have vanishing response coefficient $ B_{\hat\el+4} $,
and therefore vanishing Love number.

Although $ k_1 \neq 0 $ (see Eq.~\eqref{eq:DeltaLoveExpr}), the multipole $ \el = 1 $ is the unique
positive root of the factorization constant in Eq.~\eqref{eq:Eell}. Consequently, $ E_1 = 0 $, which implies
that $ D^{-}_{5}\left(D^{+}_{1}R_{1}\right) = 0 $. The kernel of the lowering operator, $ D^{-}_{5}\psi=0 $,
is solved by $ \psi\left(z\right) \propto z^{-3/2} $. Therefore, the raised solution is exactly
\begin{equation}
  R_{5}\propto z^{-3/2}\ .
  \label{eq:R5}
\end{equation}
Comparing this to Eq.~\eqref{eq:LoveasympDelta} for $ \hat\ell=5 $, we see that $ z^{-3/2} $ corresponds
purely to the source mode. Thus, the action of the raising operator on $ R_1 $ completely annihilates
the response mode, rendering the Love number for $ \el=5 $ exactly zero. By induction, 
using $D_{\hat\ell} R_{\hat\ell}=R_{\hat\ell+4}$, subsequent applications
of the raising operator will continue to map pure source modes to pure source modes, 
generating the entire
ladder of vanishing Love numbers for $ \el = 5,9,13\ldots $ 

A similar logic applies to the $ \el = 0,4,8\ldots $ ladder. Because the base case $ \el=0 $ 
already possesses
a vanishing response mode ($ \kappa_0 = 0 $), the entire ladder built upon it trivially inherits vanishing
Love numbers.

\subsection{Smeared source}\label{ssec:smeared}

For the source corresponding to~\eqref{eq:smearedSource} we use~\eqref{eq:falpha} in
Eq.~\eqref{eq:fullstaticperturbationeq}. Now we use the dimensionless variable
\begin{equation}
  z = \frac{r_0}{r}\ ,
  \label{eq:zsmeared}
\end{equation}
in terms of which 
\begin{equation}
f=1-z^4(1-2/z)^2=-(z-1)^2[(z-1)^2-2]\;,
\end{equation}
and the differential equation corresponding to the radial profile of the perturbations does not belong to any
particular class that has well-known solutions:
\begin{equation}
  \begin{split}
    \!\!\!
    -z^2(z-1)^2[(z-1)^2-2]R^{\prime\prime}
   % z^{2}\left(1-4z^{2}+4z^{3}-z^{4}\right)R^{\prime\prime}
     & %+ z\left(2-20z^{2}+26z^{3}-8z^{4}\right)R^{\prime}                            \\
     -2z(z-1)(4z^3-9z^2+z+1)R^\prime     \\    
     & \ \ \ + \left[-\ell\left(\ell+1\right)%-12z^{2}+18z^{3}-6z^{4}\right]R = 0 \ .
     -6z^2(z-2)(z-1)\right]R=0\ .
  \end{split}
  \label{eq:LoveEDOSmeared}
\end{equation}
This equation has five singular points, $ 0,1,1\pm\sqrt{2} $ and $ \infty $, so attempting to find an
analytical solution is hopeless. However, we can apply numerical 
methods to compute the Love numbers, as the
interior of the interval of interest, $ 0\leq z \leq 1 $ (so $r\in (r_0,+\infty)$), 
does not contain any singular points.

At spatial infinity ($ z \to 0 $), the roots of the indicial equation are $ p_{+} = \ell $ (the response mode)
and $ p_{-} = -\left(\ell+1\right) $ (the source mode). Because $\Delta p_{\ell} = \left|p_{+}-p_{-}\right| =
2\ell+1$ is always a positive integer, one has to modify the asymptotic 
solution by introducing a logarithm, to
ensure linear independence between modes,
\begin{equation}
  R^{\rm source}_{\el} = z^{-\el-1}\sum_{j=0}^{2\el}a_j z^{j} +
  c R^{\rm response} \ln \left( z \right) 
  ,
  \label{eq:RLoveResponse}
\end{equation}
otherwise the separation between solutions becomes ambiguous. 
The logarithm in~\eqref{eq:RLoveResponse} leads
to the running of the Love number, as one has to introduce an arbitrary scale $ r\sim R $ to extract a finite
coefficient from the ratio between response and source. In many contexts it is possible to relate this scale
to the renormalization group flow using effective field theory description of the black hole,
\eg,~\cite{Kol:2011vg,Barbosa:2025uau}.

At the horizon we shift the coordinate $ x = 1 - z $, such that in this 
new coordinate we apply the Frobenius
method again. The indicial equation is now $ \left(s-1\right)s+P_{0}s+Q_{0}=0 $, 
where $ P_0 =3$ and $ Q_0 
=-\ell(\ell+1)/2$ are
associated with numerical coefficients of the equation. Solutions to the indicial equation in this case are $
s_{\pm}=-1\pm\frac{1}{2}\sqrt{4+2\ell\left(\ell+1\right)} $. Noting that the discriminant is always
positive, we discard the divergent $ s_- $ branch. Near the regular solution behaves as
\begin{align}
  \begin{aligned}
    R\left(z\right)          & \simeq \left(1-z\right)^{s_{+}}\ ,         \\
    R^{\prime}\left(z\right) & \simeq -s_{+}\left(1-z\right)^{s_{+}-1}\ .
    \label{eq:LoveHorCond}
  \end{aligned}
\end{align}

The details about the numerical implementation are described in the 
Appendix~\ref{Ap:Lovenumerics}. The results obtained for the
Love numbers are parametrized as 
\begin{equation}
  k_{\el}\left( R \right) = k_{\el}^{\rm num} + \delta_{\ell}\log\left( r_0/R \right)
  \ ,
  \label{eq:ParamLoveNum}
\end{equation}
where $ R $ is a fixed distance from $ r_0 $, $ k_{\el}^{\rm num} $ is the ratio of the coefficients that is
independent of the Log branch, while $ \delta k_{\ell} $ captures the numerical coefficient that runs with the
distance from the source. We show the first few values for $ \el $ Table~\ref{tab:main_love_numbers}.
\begin{table}[htbp]
  \centering
  \renewcommand{\arraystretch}{1.6}
  \begin{tabular}{c S[table-format=4.5] c}
    \hline\hline
    Multipole ($\ell$) & {Response Ratio ($k_\ell^{\text{num}}$)} & Scale Dependence ($\delta_{ \ell }$) \\
    \hline
    0                  & 0.                                       & 0                                    \\
    1                  & -1.122                                   & 3.333                                \\
    2                  & -9.32                                    & -4.8                                 \\
    3                  & 9.857                                    & 5.229                                \\
    4                  & -28.077                                  & -1.593                               \\
    5                  & 4.458                                    & -10.217                              \\
    \hline\hline
  \end{tabular}
  \caption{Tidal response ratios and their associated scale dependence for multipoles $\ell = 0$ through
  $5$.}
  \label{tab:main_love_numbers}
\end{table}

\section{Dynamic perturbations}\label{Sec:QNMs}

In Sect.~\ref{Sec:model}, we argued that the point-like source model lacks an infinite throat at $ r_0 $. This
is confirmed by inspecting the tortoise coordinate~\eqref{eq:tortoisedef}. Using the near-horizon behavior of
$ f_0 $ in Eq.~\eqref{eq:f0nearr0}, one finds that $ r_* $ remains strictly finite as $ r \to r_0 $. This
aligns with the findings in~\cite{Nastase:2007ut}, which showed that the wave velocity is finite at the
horizon, leading to a finite time delay for perturbations to reach $ r_0 $ from the perspective of an
asymptotic observer. Because this system is conservative, purely ingoing (dissipative) boundary conditions
cannot be imposed. For this reason, dynamic perturbations in this background geometry are trivial, consisting
purely of perfectly reflected waves.

In the case of the smeared source, the fine-tuning of the couplings~\eqref{eq:couplings} is such that $
r_*\to -\infty $, and the background supports dissipative boundary conditions. Consequently, the dynamic
perturbations in this geometry have non-trivial modes and are worth investigating. In the remainder
of this section, we focus exclusively on this model, and drop the subscript $ \alpha $ from $ f_{\alpha} $
working solely with the function given by Eq.~\eqref{eq:falpha}. For dynamic perturbations $ \delta\Phi $, the
massless scalar field equation in the background~\eqref{eq:acousticDBImetric} reads
\begin{equation}
  f\partial_{r}^{2}R
  +\frac{3}{2}f^{\prime}\partial_{r}R
  -\left[\partial_{t}^{2}+\frac{3}{2}\frac{f^{\prime}}{r}+\frac{\ell\left(\ell+1\right)}{r^{2}}\right]R=0
  ,
  \label{eq:scalareomfa}
\end{equation}
where we use the ansatz $ \delta\Phi=\frac{R\left(r,t\right)}{r}Y_{\ell m}\left(\theta,\phi\right) $. Using
the tortoise coordinate, Eq.~\eqref{eq:tortoisedef}, and manipulating Eq.~\eqref{eq:scalareomfa}, we
obtain the Schrodinger-like form
\begin{equation}
  h^{\prime\prime}\left(r_{*}\right)+\left[\omega^{2}-V\left(r\right)\right]h\left(r_{*}\right)=0
  ,
  \label{eq:QNMgeneral}
\end{equation}
where $ h\left( r_* \right) $ is related to $ R $ in~\eqref{eq:scalareomfa} via transformations defined in the
Appendix~\ref{Ap:dynPert}. The potential is
\begin{equation}
  V\left(r\right)=
  \frac{3}{2}\frac{1}{r}\frac{df}{dr}+\frac{1}{2}\frac{d^{2}f}{dr^{2}}+\frac{\ell\left(\ell+1\right)}{r^{2}}
  .
  \label{eq:potF}
\end{equation}
Because we are going to solve the equation numerically, we write the potential in the implicit coordinate
$ r = r\left( r_{*} \right) $.

The initial conditions applied to~\eqref{eq:QNMgeneral} are the standard 
ingoing at the horizon and outgoing
at infinity. However, this system has a potential barrier at $ r_0 $, 
which requires some attention. In general, we
have the asymptotic forms for the solutions
\begin{equation}
  h_{\pm} \approx A_{\rm in}^{\left( \pm \right)} e^{-i k r_{*}} + A_{\rm out}^{\left( \pm \right)} e^{i k r_{*}}
  \ ,
  \label{eq:QNMaSols}
\end{equation}
where $ \pm $ indicates the sign of the infinities. At the extremal values of $r$, 
the potential~\eqref{eq:potF} behaves as
\begin{align}
  \begin{aligned}
    V\left( r\to r_0 \right) =    & \frac{2 + \el\left( \el+1 \right)}{r_0^{2}}\ , \\
    V\left( r\to \infty \right) = & 0\ .
    \label{eq:potExt}
  \end{aligned}
\end{align}
Using~\eqref{eq:QNMaSols} and the dispersion relation $ k^{2} = \omega^{2} - 
V\left( r \right) $ we have the
boundary conditions for~\eqref{eq:QNMgeneral} given by
\begin{align}
  h_{-} = & A_{\rm in}^{\left( - \right)} e^{-i \left( \sqrt{\omega^{2} - V_0} \right) r_{*}}\ , \label{eq:QNMing} \\
  h_{+} = & A_{\rm out}^{\left( + \right)} e^{i \omega r_{*}}\ , \label{eq:QNMout}
\end{align}
where $ V\left( r_0 \right) = V_0 $. In this particular case the amplitudes $ A^{\pm} $ are arbitrary, given
that Eq.~\eqref{eq:QNMgeneral} is linear and homogeneous.

As explained in the Appendix~\ref{Ap:dynPert}, the geometry 
induced by the DBI model with the smeared source
leads to a non-linear equation for the frequency $ \omega $. We adapt existing numerical routines that
transform the differential equation into an algebraic matrix problem, and find the frequencies satisfying the
system by scanning the complex plane. In Table~\ref{tab:main_QNMs} 
we show the fundamental mode for the first
few values of $ \el $. A detailed discussion of the methodology 
applied to obtain the spectrum is given in the
Appendix~\ref{Ap:dynPert}. The frequencies are parametrized as
\begin{equation}
  \omega = \omega_{\rm R} - i \omega_{\rm I}
  .
  \label{eq:QNMpar}
\end{equation}
\begin{table}[htbp]
  \centering
  \renewcommand{\arraystretch}{1.6}
  \begin{tabular}{c c c c c c}
    \hline\hline
    $\ell$ & $\sqrt{V_0}$ & $\omega_{\rm R}$ & $\omega_{\rm I}$ & $k_{\rm R}$ & $k_{\rm I}$ \\
    \hline
    0      & 1.41421      & \text{---}       & \text{---}       & \text{---}  & \text{---}  \\
    1      & 2.00000      & 0.777271         & 0.204970         & 0.085833    & -1.85613   \\
    2      & 2.82843      & 1.098708         & 0.604457         & 0.247173    & -2.68688   \\
    3      & 3.74166      & 1.771987         & 0.168780         & 0.090601    & -3.30102   \\
    4      & 4.69042      & 2.092071         & 0.494465         & 0.244317    & -4.23408   \\
    \hline\hline
  \end{tabular}
  \caption{Quasi-normal modes for various $ \el $.}
  \label{tab:main_QNMs}
\end{table}

The analogue DBI geometry is unique due to the potential barrier that naturally appears at $ r_0 $. The
effective potential in~\eqref{eq:QNMgeneral} is explicitly
\begin{equation}
  V\left( r \right) = 6\frac{r_0^{3}}{r^{5}} - 4\frac{r_0^{4}}{r^{6}} + \frac{\el\left( \el+1 \right)}{r^{2}}
  ,
  \label{eq:effPotential}
\end{equation}
which has the unusual property of being strictly monotonically decreasing, 
$ V'\left( r \right)<0 $, for $ r>r_0
$, and $ V\left( \infty \right)=0 $.

In standard black hole perturbation theory, QNMs arise from waves trapped 
within a potential well or behind a
centrifugal barrier (\eg, photon sphere). Because~\eqref{eq:effPotential} lacks any local extrema, it cannot
form a resonant cavity, and therefore this geometry cannot support standard QNM ringing.

How, then, do we interpret the complex frequencies in Table~\ref{tab:main_QNMs}? 
Observe that the real parts
of the frequencies are strictly below the potential barrier, $ \omega_R^{2}<V_0 $. 
At the horizon the momentum
$ k = \sqrt{\omega^{2} - V_0} $ introduces a branch cut in the complex frequency plane, 
starting at the points
associated with the potential barrier $ \omega = \pm \sqrt{V_0} $. 
Table~\ref{tab:main_QNMs} does not contain
the QNMs, then, but rather sub-threshold resonances, or virtual states 
(for the formal treatment see e.g., Ch.
12 of~\cite{Taylor:1972pty}).

Because the energy of these modes is below the potential barrier, the wavenumber $ k $ acquires a large
negative imaginary component ($k_I < 0$). The ingoing boundary condition $ h_- \sim e^{-i k r_*} $ thus
behaves as $ e^{k_I r_*} $. At $ r_*\to-\infty $ this wave grows exponentially, indicating that the mode is
reflected off the barrier at $ r_0 $ (formally, a pole of the S-matrix). Therefore, the horizon acts as a
reflecting wall for frequencies $ \omega^{2} < V_0 $, and the frequencies in 
Table~\ref{tab:main_QNMs} are
complex poles of purely reflected, evanescent waves leaking back to spatial infinity.

For $ \omega^{2}>V_0 $ the wave number is predominantly real, and the waves 
propagate directly into the
singularity. This is confirmed by our numerical method as we scan the complex 
plane for $ \omega_R^{2}>V_0 $
and find the complete absence of resonant poles. This is guaranteed by the smoothness of the effective
potential. Modes with $ \omega_R^{2} > V_0 $ are highly sensitive to the continuity of the potential.
From~\eqref{eq:effPotential}, one can immediately check that it is an infinitely differentiable, analytic
function. In semi-classical (WKB) scattering theory, above the barrier 
reflection for a monotonic potential is
exponentially suppressed~\cite{Landau1981Quantum}. 
As high-frequency waves adjust to the potential cliff, the
reflection coefficient vanishes exponentially, preventing any backscattering. 
Therefore, any perturbation with
frequency above $ \sqrt{V_0} $ is perfectly absorbed, precluding the 
formation of resonant quasi-normal modes.
Thus, the analogue DBI geometry acts as a classical high-pass filter, perfectly reflecting low frequency
modes, and perfectly absorbing the high frequency perturbations without ringing.

Finally, the absence of poles for $ \el = 0 $ is entirely due to the asymptotic behavior of $ V\left( r
\right) $ as $ r\to \infty $. For this monopole mode, the potential is strictly positive, and decays fast
enough to satisfy the short range integrability condition $ \int r^{2}dr V < \infty $. Under these conditions
it is proven (for e.g., Ch. 12 of~\cite{Newton:1982qc}) that the system 
cannot support any states (standard or
virtual). For $ \el \geq 1 $ the long-range centrifugal barrier breaks this integrability condition,
fundamentally changing the asymptotic solutions and allowing the existence of 
complex poles in $ \omega $.
Our numerical method captures this distinction without any extra constraints.

\section{Conclusions}

We investigated the static and dynamic perturbations around the catenoid-like 
solutions of the scalar DBI action
plus source terms, that were previously described as black hole analogues. As
shown in~\cite{Nastase:2007ut}, these perturbations obey the equation of a 
massless scalar field in a curved
background. This curved background can be described as a black hole-like 
geometry, allowing us to borrow the
methodology employed in that context to study these perturbations. 
We have considered two different sources
for the scalar DBI model, a point-like (delta) source, for the true catenoid, 
which however is not a perfect black
hole analogue as it has an infinite temperature and finite speed of propagation at the horizon;
and a distributed source $ \propto r^{2} $, shown in \cite{Nastase:2007ut} 
to be the unique scalar analogue of 
a black hole, with finite temperature and vanishing speed of propagation of information at the horizon. 
Thus the two cases differ 
by the nature of the horizon $ r_0 $. Specifically, the point-like source does not lead to an infinite
throat as one approaches $ r_0 $, such that purely ingoing boundary conditions cannot be defined. 
This also
has implications on the thermal nature of the associated black hole geometry. 
The smeared source is introduced
in order to fix these issues, and we have demonstrated that it is in fact 
possible to define purely ingoing boundary
conditions at $ r_0 $.

The static perturbations revealed two completely different behaviors 
depending on the source model. These
differences can be traced back to the scaling symmetries induced 
by the effective geometry. For the point-like
source, the effective potential exhibits the $ r^{-4} $ falloff, restricting the series recurrence relation to
steps of $ \Delta n = 4 $. Because the difference in scaling between 
source and response modes is always an
odd integer $ \Delta p = 2\el +1 $, the modes are entirely decoupled, 
yielding constant Love numbers and a
ladder symmetry analogous to the canonical acoustic black 
hole~\cite{DeLuca:2024nih}. In the case of the
smeared source the recurrence step-size is reduced to $ \Delta n = 1 $ (the choice of the dimensionless
variable $ z $ for each source model reflect the symmetry, or the lack of it). This inevitably leads to mixing
of source and response modes, and introducing a logarithmic 
branch is mathematically necessary, leading to the
running of the Love number in this case. Although the smeared 
source geometry leads to a technically more
complicated problem, we can still implement numerical 
methods to extract the numerical values associated with
the Love numbers.

The dynamic perturbations, on the other hand, are studied 
only for the smeared source model. We apply a
similar methodology to what is employed in the computation of 
quasi-normal modes, meaning that the space-time
is dissipative due to ingoing boundary condition at $ r_0 $ 
and outgoing at spatial infinity. The effective
potential associated to this geometry is a monotonically 
decreasing function from the horizon to infinity, and
has a potential barrier at the horizon. We explicitly compute the complex frequencies that solve the
eigenvalue problem, and show that they are all below the threshold energy of the potential barrier. Explicit
momentum computation reveal the evanescent character of 
modes with frequencies below the potential barrier. On
the other hand, for frequencies above the threshold there is no ringing, meaning that waves are perfectly
absorbed. This system, therefore, is analogue to a classical high-pass filter.

It is worth emphasizing that the study of perturbations on the curved geometry is far from exhausted,
especially the one related to the smeared source. The fact that this system has a thermal horizon, and the
effective potential does not contain local minima or maxima 
makes it a very interesting case to explore how
radiation propagates from the source towards infinity. 
Another intriguing aspect of this system is how far the
analogy with the analogue fluid system goes, since the map~\eqref{eq:cs}-~\eqref{eq:v0} provides a direct
relation between the solutions of the scalar DBI model and fluid variables. These points will be left for
future investigation.

%\newpage
\section*{Acknowledgments}
%%%%%%%%%%%%%%%%%%%%%%%%%%%%%%%%%%%%%%%%%%%%%%%%%%%%%%%%%%%%%%%%%%%%%%%%%%%%%%%%%%%%%%%%

We would like to thank Justin Khoury for comments.
The work of HN is supported in part by  CNPq grant 
304583/2023-5 and FAPESP grant  2024/15298-0.
HN would also like to thank the ICTP-SAIFR for their support through FAPESP grant 2021/14335-0.
The work of PM is supported by FAPESP grant 2022/12401-9.

\appendix

\section{Love numbers: numerical method used in Sect. 3.2} \label{Ap:Lovenumerics}

Eq.~\eqref{eq:LoveEDOSmeared} cannot be cast in the form of a 
Hypergeometric equation because of the singular
point structures. One can easily verify that it possesses five regular singular points, rendering a
generalized Heun differential equation. In the specific interval of interest, $ 0 \leq z \leq 1 $, which maps
the boundaries to the horizon and spatial infinity, respectively, the equation is regular and, the boundary
conditions are exactly two of the five singular points. In this interval we can use series expansion methods
and numerical integration to extract the coefficients associated with Love numbers.

The practical implementation of the code was done using \textit{Wolfram Mathematica}, 
and consists of the
following steps
\begin{itemize}
  \item Input equation~\eqref{eq:LoveEDOSmeared}, and expand $ R $ 
  in the variable $ x = 1-z $, appropriate
    for the near horizon boundary condition. Regularity is enforced by discarding the divergent branch and
    evaluate this series at a small cutoff $ z = 1 - \epsilon $ (\eg, $ \epsilon=10^{-4} $) to obtain the
    initial conditions $ R $ and $ R^{\prime} $.
  \item We integrate the differential equation outward from the horizon cutoff to the asymptotic region,
    $ z_{\rm end} = 0.1 $. Because the source mode diverges as $ z^{-\ell - 1} $ 
    and the response scales as
    $ z^{\el} $, extracting the response requires keeping track of cancellations of order $ \mathcal O\left(
    z^{-\left( 2\el +1 \right)} \right) $. At this step it is necessary to use arbitrary machine precision,
    with working precision dynamically scaled as a function of $ \el $.
  \item At spatial infinity $ z\to 0 $, we algebraically construct the exact Frobenius series up to order
    $ \mathcal O\left( z^{2\el + 15} \right) $. We first solve the recurrence relations for the pure power
    series of the decaying mode. We then construct the growing mode, explicitly injecting the logarithmic
    branch $ cR_{\rm response}\ln z $.
  \item The growing mode is defined only up to an arbitrary addition 
  of the homogeneous decaying mode. To
    uniquely fix this basis ambiguity (gauge freedom), we strictly set the coefficient of $ z^{\el} $ in the
    polynomial part of the source mode to zero. This mathematically isolates the logarithmic running
    coefficient, $ \delta_{\el} $.
  \item We evaluate the asymptotic series at the matching point 
  $ z_{\rm end} $ and match it to the numerical
    solution and its derivative. This yields a linear system that is solved to extract the source and response
    coefficients.
\end{itemize}

The convergence and stability of the routine are confirmed by varying the endpoints ($ z_{\rm end} $ and
$ \epsilon $), which are the cutoffs near spatial infinity and the horizon. The results are invariant under
these changes, and the choices used in the execution reflect 
the minimal values that reflect this invariance
to preserve performance.

\section{Dynamic perturbations} \label{Ap:dynPert}
\subsection{Reduction to Schrodinger-like equation}

Recall the equation (\ref{eq:scalareomfa}),
\begin{equation}
  f\partial_{r}^{2}R
  +\frac{3}{2}f^{\prime}\partial_{r}R
  -\left[
    \partial_{t}^{2}+\frac{3}{2}\frac{f^{\prime}}{r}+\frac{\ell\left(\ell+1\right)}{r^{2}}
  \right]R
  =
  0
  .
  \label{eq:appSchrlike}
\end{equation}
We separate the radial and time dependent parts as $ R\left(t,r\right)=e^{-i\omega t}\psi\left(r\right) $.
In the tortoise coordinate we have $ \frac{d}{dr}=f^{-1/2}\frac{d}{dr_{*}} $. Finally, apply the
transformation $ \psi\to f^{-1/2}h $ to remove the first derivative term, resulting in the equation
\begin{equation}
  \frac{d^{2}h}{dr_*^2}
  +\left[
    \omega^{2}
    -\frac{3}{2}\frac{f^{-1/2}}{r\left( r_* \right)}\frac{df}{dr_*}
    -\frac{1}{2}f^{-1}\frac{d^{2}f}{dr_*^2}
    +\frac{1}{4}f^{-2}\left(\frac{df}{dr_*}\right)^{2}
    -\frac{\ell\left(\ell+1\right)}{r^{2}\left( r_* \right)}
  \right]h
  =
  0
  ,
  \label{eq:appSchrlikeFinal}
\end{equation}
which is the equation~\eqref{eq:QNMgeneral}.

\subsection{Calculation of the complex frequencies}

\subsubsection*{Analytic manipulations}
Before the numerical implementation we need to perform some 
transformations, to turn the differential equation
into a form that is suitable for matrix discretization. Just by substitution of the ansatz
$ \delta\Phi=\frac{R\left(t,r\right)}{r}Y_{\ell m}\left(\theta,\phi\right) $ in Eq.~\eqref{eq:QNMgeneral},
with $ R\left(t,r\right)=e^{-i\omega t}R\left(r\right) $ we have
\begin{equation}
  fR^{\prime\prime}
  +\frac{3}{2}f^{\prime}R^{\prime}
  +
  \left[
    \omega^{2}-\frac{3}{2}\frac{f^{\prime}}{r}-\frac{\ell\left(\ell+1\right)}{r^{2}}
  \right]
  R=0.
  \label{eq:AppDyn1}
\end{equation}
Now introduce the coordinate
\begin{equation}
  z=1-\frac{2r_{0}}{r}
  ,
  \label{eq:AppdynZ}
\end{equation}
mapping the interval to $ -1\leq z \leq 1 $. In the $ z $ coordinate, the function $ f $ reads
\begin{equation}
  f=\frac{1}{16}\left(1+z\right)^{2}\left(7-2z-z^{2}\right)
  ,
\end{equation}
In this new coordinate the equation becomes
\begin{equation}
  \begin{split}
    \left(1-z\right)^{2}fR_{zz} + & \left[-2f+\frac{3}{2}\left(1-z\right)f_{z}\right] \left(1-z\right)R_{z} \\
                                  & {}+ \left[ \left(\frac{2r_{0}}{1-z}\right)^{2}\omega^{2}
                                  -\frac{3}{2}\left(1-z\right)f_{z}-\ell\left(\ell+1\right) \right] R=0 .
  \end{split}
\end{equation}
Next, we establish the boundary conditions in this variable. 
At $ z\to -1 $ ($r\to r_0$), we define the new variable $
y=1+z $, and the equation reads
\begin{equation}
  2y^{2}R_{yy}+6yR_{y}+\left[r^{2}_{0}\omega^{2}-\ell\left(\ell+1\right)\right]R=0
  ,
\end{equation}
in this form, we can use the ansatz $ R\sim y^{\alpha} $ yielding
\begin{equation}
  \alpha_{\pm}=-1\pm\sqrt{1-\frac{1}{2}\left[r^{2}_{0}\omega^{2}-\ell\left(\ell+1\right)\right]}
  .
\end{equation}
The ingoing behavior is associated with $ \alpha_{-} $.

At $ z\to 1 $ ($r\to \infty$, so $dr\simeq dr_*$, thus $r\simeq r_*$), 
the relation $ r_{*}\approx\frac{2r_{0}}{1-z} $ holds, and the boundary condition is
\begin{equation}
  R\sim e^{i\omega\frac{2r_{0}}{1-z}}
  .
\end{equation}

To have a smooth equation suitable for matrix discretization, 
we use the following ansatz, which contains the
boundary conditions, leading to a smooth equation for $ U\left( z \right) $:
\begin{equation}
  R=e^{i\omega\frac{2r_{0}}{1-z}}\left(1+z\right)^{\alpha_{-}}U\left(z\right)
  .
\end{equation}
The equation to be solved numerically is
\begin{equation}
  AU^{\prime\prime}+BU^{\prime}+CU=0
  ,
  \label{eq:appNumericalGen}
\end{equation}
with
\begin{equation}
  A=-\frac{1}{16}(z-1)^{2}(z+1)%\left(z^{3}+3z^{2}-5z-7\right)
^2(z^2+2z-7)  .
\end{equation}

\begin{align}
  B & =\frac{1}{16}(z+1)\left\{ -4ir_{0}\omega\left(z^{3}+3z^{2}-5z-7\right)+\left(z-1\right)
  \left[10+7\sqrt{2}\sqrt{2+\ell\left(\ell+1\right)-r^{2}_{0}\omega^{2}}\right.\right.\nonumber \\
    & +\left(22-9\sqrt{2}\sqrt{2+\ell\left(\ell+1\right)-r^{2}_{0}\omega^{2}}\right)z+\left(-10+\sqrt{2}\sqrt{2+
    \ell\left(\ell+1\right)-r^{2}_{0}\omega^{2}}\right)z^{2}\nonumber \\
    & +\left.\left.\left(-6+\sqrt{2}\sqrt{2+\ell\left(\ell+1\right)-r^{2}_{0}\omega^{2}}\right)z^{3}\right]\right\}
    .
\end{align}

\begin{align}
  C & =\frac{1}{32}\left(1+z\right)\left\{ 4ir_{0}\omega\left[4-7\sqrt{2}
  \sqrt{2+\ell\left(\ell+1\right)-r^{2}_{0}\omega^{2}}\right.\right.\nonumber \\
    &\left.+\left(-4+\sqrt{2}\sqrt{2+\ell\left(\ell+1\right)-r^{2}_{0}\omega^{2}}\right)
    z\left(2+z\right)\right]\nonumber \\
    & +\left(z-1\right)\left[2-17\sqrt{2}\sqrt{2+\ell\left(\ell+1\right)-r^{2}_{0}\omega^{2}}\right.\nonumber \\
    &+4\left(-4+\sqrt{2}\sqrt{2+\ell\left(\ell+1\right)-r^{2}_{0}\omega^{2}}\right)z\nonumber \\
    &\left.+\left(-2+5\sqrt{2}\sqrt{2+\ell\left(\ell+1\right)-r^{2}_{0}\omega^{2}}
    \right)z^{2}\right]\nonumber                                             \\
    & +\left.\ell\left(\ell+1\right)\left(-z^{3}+z^{2}+9z-25\right)+r^{2}_{0}
    \omega^{2}\left(z^{3}-z^{2}-z+65\right)\right\}
    .
\end{align}

\subsubsection*{Numerical implementation}
The potential barrier introduces a non-analytic dependence on the 
frequency via the asymptotic momentum $ k =
\sqrt{\omega^{2} - V_0} $, so the perturbation equation cannot be linearized into a standard generalized
eigenvalue problem. To preserve the non-linear structure of the boundary conditions without resorting to
low-frequency approximations, we treat the system as a non-linear eigenvalue problem.

The procedure outlined below is inspired by~\cite{Jansen:2017oag}, 
which suggests the solution of an equation
of the type $ \det M = 0 $ while scanning the complex plane for $ \omega $. 
While this method preserves the
non-linear dependence of $ \omega $, finite matrix truncations in the presence 
of branch-cuts are known to
generate spectral pollution — spurious, non-convergent numerical roots~\cite{boyd01}. 
To distinguish between
numerical artifacts and true resonances we employ the criterion of convergence 
at spatial infinity established
in~\cite{Leaver:1985ax}.

The matrix discretization is obtained by mapping the coordinate $ y = z+1 $ and expanding
\begin{equation}
  U\left(y\right)=\sum^{N}_{n=0}a_{n}y^{n}
  \label{eq:appSeriesU}
  .
\end{equation}
The differential equation has strictly polynomial coefficients, so the series yields a band-diagonal
recurrence relation for $ a_n $'s,
\begin{equation}
  \sum_{j}M_{nj}\left(\omega,\ell\right)a_{j}=0
  ,
\end{equation}
where the matrix coefficients $ M_{nj} $ depend on $ \omega $, $ \el $ and on the grid index $ n $.

We look for complex frequencies $ \omega = \omega_R - i \omega_I $ that satisfy the condition
\begin{equation}
  \det\left(M\right)=0
  \label{eq:appDet}
  .
\end{equation}
These roots are obtained using arbitrary-precision secant method (method of principal axis) with 40 digit
precision to bypass the floating-point noise floor. Specifically, high-precision arithmetic is required
because evaluating the boundary convergence at spatial infinity for large $ N $ exponentially amplifies
numerical truncation errors, which would falsely reject true resonant poles 
if standard machine precision were used.

To distinguish the actual frequencies that solve~\eqref{eq:appDet} 
from numerical artifacts, we evaluate the
convergence of the series~\eqref{eq:appSeriesU} at spatial infinity $ y=2 $. If $ \omega $ is a solution to
the determinant equation, then $ U\left( y \right) $ is analytically smooth at infinity and the series
converges super-exponentially. On the other hand, if $ \omega $ 
approximates a branch-cut the series becomes
singular at the boundary, limiting the radius of convergence to $ R=2 $.

To assess this information we define the boundary convergence error 
$ \epsilon_N $ as the magnitude of the
$ N $th term in the series~\eqref{eq:appSeriesU} at infinity
\begin{equation}
  \epsilon_{N}=\left|a_{N}2^{N}\right|
  .
\end{equation}
We classify the frequency as a true solution if $ \epsilon_N $ exhibits exponential decays with increasing
$ N $ ($ \epsilon_N \ll 1 $ and strictly decreasing). Frequencies for which $ \epsilon_N\sim \cal O\left( 1
\right) $ are discarded as artifacts.

To validate this approach we employ some sanity checks. First, we obtain QNMs associated with the
Schwarzschild black hole, which are well documented and computed to very high precision, for instance
in~\cite{Berti:2009kk}. Our numerical routine reproduces the 
results for the scalar QNMs in that paper up to 6
decimal places, thus, we assert that it is a reliable method for extracting QNMs. 
On top of the analyticity test
for rejecting pseudo-poles, we test the frequencies obtained for invariance under grid scaling, running the
numerical routine multiple times for different $N$. In doing so, we see that frequencies associated with
branch-cuts exhibit high deviation in between different runs, whereas 
QNM frequencies persist up to several
(more than 8) decimal digits. The numerical routines employed in the calculation of the values reported in
Table~\ref{tab:main_QNMs}, as well as check against known values for 
QNMs are available at  https://github.com/pmeert/scalarDBIperturbation .

\bibliographystyle{utphys}
\bibliography{DBIbibliography}
\end{document}